\documentclass[preprint,preprintnumbers,amsmath,amssymb]{revtex4}
\usepackage{amsmath}
\usepackage{bm}
\usepackage{epsfig}
\date{}
\begin{document}
\title{Collision of Multimode Dromions and a Firewall in the Two Component Long Wave Short Wave
Resonance Interaction Equation}
\author{R. Radha$^1$, C. Senthil Kumar$^2$, M. Lakshmanan$^3$, C. R. Gilson$^4$}
  \address{
  $^1$ Centre for Nonlinear Science, Dept. of Physics, Govt. College for Women, Kumbakonam 612 001, India \\
  $^2$ Dept. of Physics, VMKV Engineering College,
  Periaseeragapadi, Salem - 636 308, India \\
  $^3$ Centre for Nonlinear Dynamics, Dept. of Physics, Bharathidasan University, Tiruchirapalli-620 024, India\\
  $^4$ Dept. of Mathematics, University of Glasgow, Glasgow, UK.}

\begin{abstract}
In this paper, we investigate the two component long wave short
wave resonance interaction (2CLSRI) equation and show that it
admits the Painleve property. We then suitably exploit the
recently developed truncated Painleve approach to generate
exponentially localized solutions for the short wave components
$S^{(1)}$ and $S^{(2)}$ while the long wave L admits line soliton
only. The exponentially localized solutions driving the short
waves $S^{(1)}$ and $S^{(2)}$ in the y direction are endowed with
different energies (intensities) and are called "multimode
dromions". We also observe that the multimode dromions suffer
intramodal inelastic collision while the existence of a firewall
across the modes prevents the switching of energy between the
modes.
\end{abstract}

\maketitle

\section{Introduction}
Recent investigations of the integrable coupled nonlinear
Schr\"odinger equation, namely the celebrated Manakov model, and
the observation of intensity redistribution in the collision of
solitons  [1-5] have clearly pointed out their potential usage in
the field of optical communications and have virtually set in
motion the process of designing an all optical computing machine.
In particular, the vector solitons undergoing energy sharing
collision identified in the coupled NLS equation turned out to be
the key in the growing list of alternatives to the paradigm of
soliton based chips, at least for specialized applications
including quantum computing [6],  DNA computing [7] and dynamics
based computing based on chaos [8].

It is known that the identification of dromions [9,10] in the
Davey-Stewartson equation which is a (2+1) dimensional
generalization of the NLS equation has given the much needed
impetus to the investigation of (2+1) dimensional integrable
models.  These dromions which are localized exponentially in all
directions are essentially driven by certain lower dimensional
arbitrary functions of space and time.  In fact, such lower
dimensional arbitrary functions of space and time have
consolidated the concept of integrability of the associated
dynamical systems in (2+1) dimensions besides being tailor made
for the construction of various kinds of localized solutions.
Reflecting on the flurry of activities taking place in the field
of optical communication ever since the identification of shape
changing collision of vector solitons in the coupled NLS equation
and the rapid strides made in the field of (2+1) dimensional
nonlinear partial differential equations (pdes) after the
observation of dromions in the Davey-Stewartson I (DSI) equations,
one would be tempted to look for the possibility of identifying
the counterparts of vector solitons in (2+1) dimensions as well.

In fact, the  recent derivation of the two component long wave
short wave resonance interaction (2CLSRI) equation in the context
of the interaction of nonlinear dispersive waves on three channels
[11] has only fuelled the anticipation to look for such localized
excitations. This is also further supported by the study of
collision behaviour of plane solitons admitted by the 2CLSRI
equation recently [12]. In this paper, we investigate 2CLSRI
equation and confirm its Painleve property. We then suitably
employ the recently developed truncated Painleve approach [13-16]
and generate multimode dromions. It should be mentioned that this
is the first time the existence of exponentially localized
solutions has been reported in a vector (2+1) dimensional
nonlinear pde. Finally, we also study the unusual interaction of
multimode dromions.

We now consider the two component long wave short wave resonance
interaction (2CLSRI) equation in the following form
\begin{subequations}
\begin{eqnarray}
i(S_{t}^{(1)}+S_{y}^{(1)})-S_{xx}^{(1)}+LS^{(1)}=0, \\
i(S_{t}^{(2)}+S_{y}^{(2)})-S_{xx}^{(2)}+LS^{(2)}=0, \\
L_{t}=2(|S^{(1)}|^2_x+|S^{(2)}|^2_x).
\end{eqnarray} \label{I}
\end{subequations}
The above equation is the two component analogue of the long wave
short wave resonance interaction equation investigated recently
[13]. In eq.(1), $S^{(1)}$ and $S^{(2)}$ represent short waves
while L denotes a long wave. In particular, it explains the
interaction of long interfacial wave (L) and a short surface wave
(S) in a two layer fluid. This equation has been investigated
recently and line solitons have been generated [11,12].

\section{Singularity Structure Analysis}
We now rewrite the above equation by putting $S^{(1)}=p$,
${S^{(1)}}^*=q$, $S^{(2)}=r$, ${S^{(2)}}^*=s$ as
\begin{subequations}
\begin{eqnarray}
i(p_{t}+p_{y})-p_{xx}+Lp=0, \\
-i(q_{t}+q_{y})-q_{xx}+Lq=0,  \\
i(r_{t}+r_{y})-r_{xx}+Lr=0, \\
-i(s_{t}+s_{y})-s_{xx}+Ls=0,  \\
L_{t}=2(pq)_{x}+2(rs)_{x}.
\end{eqnarray} \label{s1}
\end{subequations}
We now effect a local Laurent expansion of the variables $p$, $q$,
$r$, $s$ and $L$ in the neighbourhood of a noncharacteristic
singular manifold $\phi(x,y,t)=0$, $\phi_x\neq0$, $\phi_y\neq0$.
Assuming the leading order of the solutions of eq. (2) to have the
following form
\begin{equation}
p=p_{0}\phi^{\alpha},
q=q_{0}\phi^{\beta},r=r_{0}\phi^{\gamma},s=s_{0}\phi^{\delta},L=L_{0}\phi^m,
\label{s2}
\end{equation}
where $p_{0}$, $q_{0}$, $r_{0}$, $s_{0}$ and $L_{0}$ are analytic
functions of ($x$, $y$, $t$) and $\alpha$, $\beta$, $\gamma$,
$\delta$ and $m$ are integers to be determined, we now substitute
(3) into (2) and balance the most dominant terms to obtain
\begin{equation}
\alpha=\beta=\gamma=\delta=-1,m=-2,
\end{equation}
with the condition
\begin{equation}
p_0q_0+r_{0}s_{0}=\phi_{x}\phi_{t}, L_{0}=2 \phi_{x}^2. \label{s6}
\end{equation}

Now, considering the generalized Laurent expansion of the
solutions in the neighbourhood of the singular manifold
\begin{subequations}
\begin{eqnarray}
p=p_{0}\phi^{\alpha}+...+p_{j}\phi^{j+\alpha}+..., \\
q=q_{0}\phi^{\beta}+...+q_{j}\phi^{j+\beta}+..., \\
r=r_{0}\phi^{\gamma}+...+r_{j}\phi^{j+\gamma}+..., \\
s=s_{0}\phi^{\delta}+...+s_{j}\phi^{j+\delta}+..., \\
L=L_{0}\phi^{\omega}+...+L_{j}\phi^{j+\omega}+...,
\end{eqnarray}\label{s3}
\end{subequations}
the resonances which are the powers at which arbitrary functions
enter into (6) can be determined by substituting (6) into (2).
Vanishing of the coefficients of
($\phi^{j-3}$,$\phi^{j-3}$,$\phi^{j-3}$,$\phi^{j-3}$,$\phi^{j-3}$)
leads to the condition
\begin{equation}
\left(
\begin{array}{ccccc}
-j(j-3)\phi_{x}^2  & 0                 & 0                & 0                   & p_{0} \\
0                  & -j(j-3)\phi_{x}^2 & 0                & 0                   & q_{0} \\
0                  &0                  & -j(j-3)\phi_{x}^2 & 0                  & r_{0} \\
0                  &0                  &0                  & -j(j-3)\phi_{x}^2  & s_{0} \\
2(j-2)q_{0}\phi_{x}& 2(j-2)p_{0}\phi_{x}&
2(j-2)s_{0}\phi_{x}&2(j-2)r_{0}\phi_{x} & -(j-2)\phi_{t}
\end{array}
\right) \left(
\begin{array}{c}
p_{j} \\
q_{j} \\
r_{j} \\
s_{j} \\
L_{j}
\end{array}
\right) = 0. \label{s4}
\end{equation}
From equation (7), one gets the resonance values as
\begin{equation}
j=-1,\ 0,0,0,\ 2,\ 3,\ 3,\ 3,\ 4.
\end{equation}

The resonance at $j$ = -1 naturally represents the arbitrariness
of the manifold $\phi(x,y,t)=0$. In order to prove the existence
of arbitrary functions at the other resonance values, we now
substitute the full Laurent series

\begin{eqnarray}
p=p_{0}\phi^{\alpha}+\sum_{j}p_{j}\phi^{j+\alpha}, \quad
q=q_{0}\phi^{\beta}+\sum_{j}q_{j}\phi^{j+\beta}, \quad
r=r_{0}\phi^{\gamma}+\sum_{j}r_{j}\phi^{j+\gamma},\nonumber \\
s=s_{0}\phi^{\delta}+\sum_{j}s_{j}\phi^{j+\delta}, \quad
L=L_{0}\phi^{\omega}+\sum_{j}L_{j}\phi^{j+\omega}
\end{eqnarray} \label{s13}

into equation (2).  Now, collecting the coefficients of
($\phi^{-3}$,$\phi^{-3}$,$\phi^{-3}$,$\phi^{-3}$,$\phi^{-3}$) and
solving the resultant equation, we obtain equation (5), implying
the existence of a
resonance at $j=0,0,0$.\\
Similarly, collecting the  coefficients of
($\phi^{-2}$,$\phi^{-2}$,$\phi^{-2}$, $\phi^{-2}$, $\phi^{-2}$)
and solving the resultant equations by using the Kruskal's ansatz,
$\phi(x,y,t)=x+\psi(y,t)$, we get
\begin{subequations}
\begin{eqnarray}
p_{1}=\frac{1}{2}[ip_{0}(\psi_{t}+\psi_{y})-2p_{0x}],\\
q_{1}=\frac{1}{2}[-iq_{0}(\psi_{t}+\psi_{y})-2q_{0x}],\\
r_{1}=\frac{1}{2}[ir_{0}(\psi_{t}+\psi_{y})-2r_{0x}], \\
s_{1}=\frac{1}{2}[-is_{0}(\psi_{t}+\psi_{y})-2s_{0x}],\\
L_{1}=0.
\end{eqnarray}\label{s14}
\end{subequations}\\

Collecting the coefficients  of
($\phi^{-1}$,$\phi^{-1}$,$\phi^{-1}$,$\phi^{-1}$,$\phi^{-1}$), we
have
\begin{subequations}
\begin{eqnarray}
&&i(p_{0t}+p_{0y})-p_{0xx}+L_{0}p_{2}+L_{1}p_{1}+L_{2}p_{0}=0,\label{s7}\\
&&-i(q_{0t}+q_{0y})-q_{0xx}+L_{0}q_{2}+L_{1}q_{1}+L_{2}q_{0}=0,\label{s7_1} \\
&&i(r_{0t}+r_{0y})-r_{0xx}+L_{0}r_{2}+L_{1}r_{1}+L_{2}r_{0}=0,\label{s8} \\
&&-i(s_{0t}+s_{0y})-s_{0xx}+L_{0}s_{2}+L_{1}s_{1}+L_{2}s_{0}=0,\label{s8_1} \\
&&\hspace{-2cm} L_{1t}=2[p_{0x}q_{1}+q_{1x}p_{0}+p_{1x}q_{0}+p_{1}q_{0x}]+
2[r_{0x}s_{1}+r_{1x}s_{0}+s_{1x}r_{0}+r_{1}s_{0x}]=0.
\end{eqnarray}\label{s8_2}

From (11a),(11b),(11c) and (11d), we can eliminate $L_{2}$ to
obtain the following three equations for the four unknowns $p_2$,
$q_2$, $r_2$ and $s_2$,

\begin{eqnarray}
&&\hspace{-2cm}L_{0}(p_{0}q_{2}-q_{0}p_{2})-(p_{0}q_{0xx}-q_{0}p_{0xx})
-i(p_{0}(q_{0t}+q_{0y})+q_{0}(p_{0t}+p_{0y}))=0, \\
&&\hspace{-2cm} L_{0}(p_{0}r_{2}-r_{0}p_{2})-(p_{0}r_{0xx}-r_{0}p_{0xx})
-i(-p_{0}(r_{0t}+r_{0y})+r_{0}(p_{0t}+p_{0y}))=0, \\
&&\hspace{-2cm} L_{0}(p_{0}s_{2}-s_{0}p_{2})-(p_{0}s_{0xx}-s_{0}p_{0xx})
-i(p_{0}(s_{0t}+s_{0y})+s_{0}(p_{0t}+p_{0y}))=0,
\end{eqnarray}
\end{subequations}
which ensures that one of the functions $p_2$, $q_2$, $r_2$ or
$s_2$ is arbitrary. Obviously $L_2$ itself can be obtained from
any one of the four equations (11a), (11b), (11c) or (11d).
Similarly, collecting the coefficients of
($\phi^{0}$,$\phi^{0}$,$\phi^{0}$,$\phi^{0}$,$\phi^{0}$), we have

\begin{subequations}
\begin{eqnarray}
i(p_{1t}+p_{2}\psi_{t})+i(p_{1y}+p_{2}\psi_{y})-(p_{1xx}+2p_{2x})
+L_{2}p_{1}+L_{3}p_{0}=0,\label{s9} \\
-i(q_{1t}+q_{2}\psi_{t})-i(q_{1y}+q_{2}\psi_{y})-(q_{1xx}+2q_{2x})
+L_{2}q_{1}+L_{3}q_{0}=0,\label{s9_1} \\
i(r_{1t}+r_{2}\psi_{t})+i(r_{1y}+r_{2}\psi_{y})-(r_{1xx}+2r_{2x})
+L_{2}r_{1}+L_{3}r_{0}=0,\label{s10} \\
-i(s_{1t}+s_{2}\psi_{t})-i(s_{1y}+s_{2}\psi_{y})-(s_{1xx}+2s_{2x})
+L_{2}s_{1}+L_{3}s_{0}=0,\label{s10_1} \\
L_{2t}+L_{3}\psi_{t}=2[p_{0x}q_{2}+(p_{1x}+p_{2})q_{1}+(p_{2x}+p_{3})q_{0}
\nonumber\\
+q_{0x}p_{2}+(q_{1x}+q_{2})p_{1}+(q_{2x}+q_{3})p_{0}]+
\nonumber\\
2[r_{0x}s_{2}+(r_{1x}+r_{2})s_{1}+(r_{2x}+r_{3})s_{0}
\nonumber\\
+s_{0x}r_{2}+(s_{1x}+s_{2})r_{1}+(s_{2x}+s_{3})r_{0}].
\end{eqnarray}

Equations (12a), (12b), (12c) and (12d) can be solved for $L_3$ as
\begin{eqnarray}
L_3=\frac{1}{p_0}(-i(p_{1t}+p_{2}\psi_{t})-i(p_{1y}+p_{2}\psi_{y})+(p_{1xx}+2p_{2x})
-L_{2}p_{1}), \label{s11} \\
L_3=\frac{1}{q_0}(i(q_{1t}+q_{2}\psi_{t})+i(q_{1y}+q_{2}\psi_{y})+(q_{1xx}+2q_{2x})
-L_{2}q_{1}), \label{s11_1} \\
L_3=\frac{1}{r_0}(-i(r_{1t}+r_{2}\psi_{t})-i(r_{1y}+r_{2}\psi_{y})+(r_{1xx}+2r_{2x})
-L_{2}r_{1}) \label{s12_1}\\
L_3=\frac{1}{s_0}(i(s_{1t}+s_{2}\psi_{t})+i(s_{1y}+s_{2}\psi_{y})+(s_{1xx}+2s_{2x})
-L_{2}s_{1}). \label{s12}
\end{eqnarray}
\end{subequations}
Making use of eqns. (5), (10) and (11), we find that the right
hand sides of eqs. (12f), (12g), (12h) and (12i) are equal. This
implies that we are left with two equations for five unknowns. So,
any three of the five coefficients $p_3$, $q_{3}$, $r_{3}$, $s_3$
or $L_{3}$ are arbitrary. Now, collecting the coefficients of
($\phi$, $\phi$, $\phi$, $\phi$, $\phi$), we have
\begin{subequations}
\begin{eqnarray}
&&i(p_{2t}+2p_{3}\psi_{t})+i(p_{2y}+2p_{3}\psi_{y})-(p_{2xx}+4p_{3x}+6p_{4})
\nonumber \\
&&+L_{0}p_{4}+L_{2}p_{2}+L_{3}p_{1}+L_{4}p_{0}=0,\label{s13_1} \\
&&-i(q_{2t}+2q_{3}\psi_{t})-i(q_{2y}+2q_{3}\psi_{y})-(q_{2xx}+4q_{3x}+6q_{4})
\nonumber \\
&&+L_{0}q_{4}+L_{2}q_{2}+L_{3}q_{1}+L_{4}q_{0}=0, \label{s13_2}\\
&&i(r_{2t}+2r_{3}\psi_{t})+i(r_{2y}+2r_{3}\psi_{y})-(r_{2xx}+4r_{3x}+6r_{4})
\nonumber \\
&&+L_{0}r_{4}+L_{2}r_{2}+L_{3}r_{1}+L_{4}r_{0}=0, \label{s13_3}\\
&&-i(s_{2t}+2s_{3}\psi_{t})-i(s_{2y}+2s_{3}\psi_{y})-(s_{2xx}+4s_{3x}+6s_{4})
\nonumber \\
&&+L_{0}s_{4}+L_{2}s_{2}+L_{3}s_{1}+L_{4}s_{0}=0, \label{s13_4}\\
&&\hspace{-2cm} L_{3t}+2L_{4}\psi_{t}=2[p_{0x}q_{3}-p_{0}q_{4}+(p_{1x}+p_{2})q_{2} \nonumber \\
&&+(p_{2x}+2p_{3})q_{1}+(p_{3x}+3p_{4})q_{0}+q_{0x}p_{3}-q_{0}p_{4} \nonumber \\
&&+(q_{1x}+q_{2})p_{2}+(q_{2x}+2q_{3})p_{1}+(q_{3x}+3q_{4})p_{0}]+ \nonumber \\
&&2[r_{0x}s_{3}-r_{0}s_{4}+(r_{1x}+r_{2})s_{2}+(r_{2x}+2r_{3})s_{1}+\nonumber \\
&&(r_{3x}+3r_{4})s_{0}+s_{0x}r_{3}-s_{0}r_{4} \nonumber \\
&&+(s_{1x}+s_{2})r_{2}+(s_{2x}+2s_{3})r_{1}+(s_{3x}+3s_{4})r_{0}].
\label{s13_5}
\end{eqnarray}
\end{subequations}
By multiplying (13a) by $q_0$, (13b) by $p_0$, (13c) by $s_0$,
(13d) by $r_0$ and adding the resultant equation, we obtain an
equation which is same as (13e). This means that we have only four
determining equations for five unknowns.  So, any one of the five
functions $p_4$, $q_{4}$, $r_{4}$, $s_4$ or $L_{4}$ is arbitrary.
One can proceed further to determine all other coefficients of the
Laurent expansions (9) without the introduction of any movable
critical manifold. Thus, the 2CLSRI equation indeed satisfies the
Painlev\'e property.

\section{Truncated Painleve Approach and Localized Solutions}
To generate the solutions of 2CLSRI equation, we now suitably
exploit the results of the leading order behaviour by employing
the truncated Painlev\'e approach. Truncating the Laurent series
of the solutions of eq. (2) at the constant level term, one
obtains following the B\"acklund transformation

\begin{eqnarray}
p=\frac{p_0}{\phi}+p_1,\quad
q=\frac{q_0}{\phi}+q_1,\quad
r=\frac{r_0}{\phi}+r_1,\quad
s=\frac{s_0}{\phi}+s_1,\quad\nonumber\\
L=\frac{L_0}{\phi^2}+\frac{L_1}{\phi}+L_2.
\end{eqnarray} \label{p1}

Assuming the following seed solution
\begin{equation}
p_1=q_1=r_1=s_1=0, \quad L_2=L_2(x,y), \label{p2}
\end{equation}
we now substitute (14) with the above seed solution (15) into
equations (2) and obtain (5) by collecting the coefficients of
$(\phi^{-3},\phi^{-3},\phi^{-3},\phi^{-3},\phi^{-3})$. Gathering
the coefficients of
$(\phi^{-2},\phi^{-2},\phi^{-2},\phi^{-2},\phi^{-2})$, we have the
following system of equations
\begin{subequations}
\begin{eqnarray}
-ip_0\phi_t-ip_0\phi_y+2p_{0x}\phi_x+p_0\phi_{xx}+L_1p_0=0, \label{p5a} \\
iq_0\phi_t+iq_0\phi_y+2q_{0x}\phi_x+q_0\phi_{xx}+L_1q_0=0,\label{p6a} \\
-ir_0\phi_t-ir_0\phi_y+2r_{0x}\phi_x+r_0\phi_{xx}+L_1r_0=0, \label{p5} \\
is_0\phi_t+is_0\phi_y+2s_{0x}\phi_x+s_0\phi_{xx}+L_1s_0=0,\label{p6} \\
L_{0t}-L_1 \phi_t = 2 (p_0q_0+r_0s_0)_x. \label{p3}
\end{eqnarray} \label{p7}
\end{subequations}
From equation (16e), we have
\begin{equation}
L_1=2\frac{(\phi_x \phi_{tx}-\phi_{xx}\phi_t)}{\phi_t}. \label{p4}
\end{equation}
Using (17) in (16a-16d), the variables $p_0$, $q_0$, $r_0$ and
$s_0$ can be solved as
\begin{subequations}
\begin{eqnarray}
&&p_0=F_1(y,t)\mbox{exp}\bigg[{\frac{1}{2}\int\frac{i(\phi_t+\phi_y)+\phi_{xx}
-\frac{2\phi_x\phi_{tx}}{\phi_t}} {\phi_{x}}{\rm d}x}\bigg],
\label{p10a}\nonumber\\\\
&&q_0=F_1(y,t)\mbox{exp}\bigg[{\frac{1}{2}\int\frac{-i(\phi_t+\phi_y)+\phi_{xx}
-\frac{2\phi_x\phi_{tx}}{\phi_t}} {\phi_{x}}{\rm d}x}\bigg],
\label{p10b}\nonumber\\\\
&&r_0=F_2(y,t)\mbox{exp}\bigg[{\frac{1}{2}\int\frac{i(\phi_t+\phi_y)+\phi_{xx}
-\frac{2\phi_x\phi_{tx}}{\phi_t}} {\phi_{x}}{\rm
d}x}\bigg],\label{p10c}\nonumber\\\\
&&s_0=F_2(y,t)\mbox{exp}\bigg[{\frac{1}{2}\int\frac{-i(\phi_t+\phi_y)+\phi_{xx}
-\frac{2\phi_x\phi_{tx}}{\phi_t}} {\phi_{x}}{\rm
d}x}\bigg],\label{p10d}\nonumber\\
\end{eqnarray}\label{p4_1}
\end{subequations}
where $F_1(y,t)$ and $F_2(y,t)$ are lower dimensional arbitrary
functions of $y$ and $t$.

Substituting (18) in (5), we obtain the condition
\begin{equation}
F_2(t-y)^2=\phi_t-F_1(t-y)^2.
\end{equation}

Again, collecting the coefficients of
$(\phi^{-1},\phi^{-1},\phi^{-1},\phi^{-1},\phi^{-1})$, we have
\begin{subequations}
\begin{eqnarray}
ip_{0t}+ip_{0y}-p_{0xx}+L_2p_0 =0, \label{p13a} \\
-iq_{0t}-iq_{0y}-q_{0xx}+L_2q_0 =0, \label{p14a} \\
ir_{0t}+ir_{0y}-r_{0xx}+L_2r_0 =0, \label{p13} \\
-is_{0t}-is_{0y}-s_{0xx}+L_2s_0 =0, \label{p14} \\
L_{1t}=0. \label{p8}
\end{eqnarray}
\end{subequations}
Making use of (17), we rewrite (20e) in the following trilinear
form
\begin{equation}
\phi_t^2\phi_{xxt}-\phi_x\phi_{tx}\phi_{tt}+\phi_{xt}^2\phi_t
+\phi_{x}\phi_{ttx}\phi_{t}=0. \label{p9}
\end{equation}
The above trilinear equation ensures that the arbitrary manifold
$\phi(x,y,t)$ should be partitioned as
\begin{equation}
\phi = \phi_1(x,y)+\phi_2 (y,t), \label{p10}
\end{equation}
where $\phi_1(x,y)$ and $\phi_2 (y,t)$ are arbitrary functions in
the indicated variables.  Making use of (22) in eqs. (18a) and
(18b), one can show that eqs. (20a-20d) are consistent provided
the submanifold $\phi_2(y,t)$ can be split as
\begin{equation}
\phi_2(y,t) = \phi_{21}(y)+\phi_{22}(t-y), \label{p16}
\end{equation}
Again, collecting the coefficients of
$(\phi^{0},\phi^{0},\phi^{0},\phi^{0},\phi^{0})$, we have only one
equation
\begin{equation}
L_{2t}=0. \label{p15}
\end{equation}
Making use of  (20a) for $L_2$, (24) reduces to the form
\begin{equation}
(F_{1tt}+F_{1ty})F_1+(F_{1t}+F_{1y}) F_{1t} =0. \label{p15_1}
\end{equation}
Equation (25) can be solved to obtain the form for $F_1(y,t)$ as
\begin{equation}
F_1(y,t)=F_1(t-y).
\end{equation}
Thus, the solutions of 2CLSRI can be written as
\begin{subequations}
\begin{eqnarray}
S^{(1)}(x,y,t)&=&\frac{F_1(t-y)\sqrt{
\phi_{1x}}e^{\int\frac{1}{2}\frac{i(\phi_{1y}+\phi_{21,y})}
{\phi_{1x}}{\rm
d}x}}{(\phi_1(x,y)+\phi_{21}(y)+\phi_{22}(t-y))},\label{p20a} \nonumber\\\\
S^{(2)}(x,y,t)&=&\frac{\sqrt{(\phi_{22,t}-F_1(t-y)^2)\phi_{1x}}
e^{\int\frac{1}{2}\frac{i(\phi_{1y}+\phi_{21,y})} {\phi_{1x}}{\rm
d}x}}{(\phi_1(x,y)+\phi_{21}(y)+\phi_{22}(t-y))},
\label{p20b}\nonumber\\
\end{eqnarray}
\end{subequations}
\begin{eqnarray}
L&=&\frac{2
\phi_{1x}^2}{(\phi_1(x,y)+\phi_{21}(y)+\phi_{22}(t-y))^2}\nonumber\\
&-& \frac{2
\phi_{1xx}}{(\phi_1(x,y)+\phi_{21}(y)+\phi_{22}(t-y))}+L_2,
\label{p19}
\end{eqnarray}
where
\begin{eqnarray}
L_2&=& \int\frac{1}{2}\bigg(
\frac{(\phi_{1yy}+\phi_{21,yy})-i\phi_{1xxy}}{\phi_{1x}}
-\nonumber\\
\quad &&\frac{(\phi_{1y}+\phi_{21,y})-i\phi_{1xx}}{\phi_{1x}^2}\phi_{1xy} \bigg)dx \nonumber\\
&+&\frac{1}{2}\frac{i\phi_{1xy}+\phi_{1xxx}}{\phi_{1x}}
-\frac{1}{4}\frac{(\phi_{1y}+\phi_{21,y})^2+\phi_{1xx}^2}{\phi_{1x}^2}.
\end{eqnarray}

Thus, by choosing the arbitrary functions $F_1(t-y)$,
$\phi_1{(x,y)}$, $\phi_{21}(y)$ and $\phi_{22}(t-y)$ suitably, one
can generate various kinds of localized solutions for the short
waves $S^{(1)}$ and $S^{(2)}$ while the longwave L does not
support completely localized solutions. From (27a) and (27b), it
is also obvious that the two physical fields $S^{(1)}$ and
$S^{(2)}$ have the same form except that their amplitudes are
different and are driven by arbitrary functions $F_1(t-y)$ and
$\sqrt{\phi_{22,t}-F_1(t-y)^2}$, respectively. It is also obvious
that the 2CLSRI equation possesses an extra arbitrary function of
space and time in comparison with its scalar counterpart [13].

\section{Dromion solutions and their Interactions}
Now we choose specific forms of the arbitrary functions in (28)
and (29) and obtain explicit exponentially localized dromion
solutions and study their interactions. To generate a (1,1)
dromion for the modes $S^{(1)}$ and $S^{(2)}$, we choose the lower
dimensional arbitrary functions of space and time, for example, as

\begin{eqnarray}
F_1(t-y) = a_1 sech(d_1 (t - y) +e_1) +g_1\nonumber\\
\phi_1(x,y) = a_2 tanh(b_2 x +c_2 y +e_2)+g_2\\
\phi_{21}(y)=g_3, \phi_{22}(t-y)=a_4 tanh (d_4 (t-y) +e_4)
+g_4\nonumber
\end{eqnarray}

\begin{figure}
\epsfig{figure=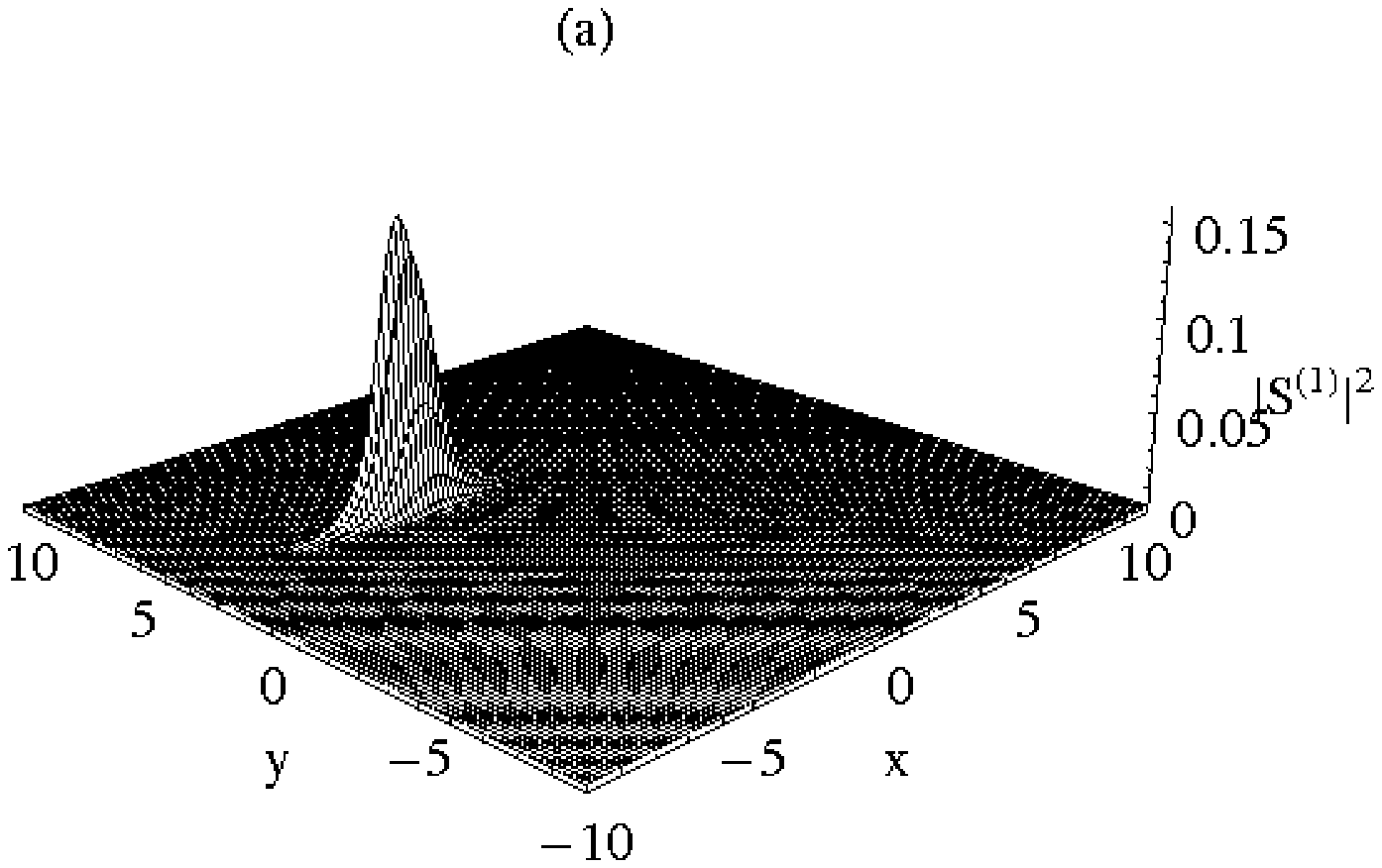, width=0.4\linewidth}
\epsfig{figure=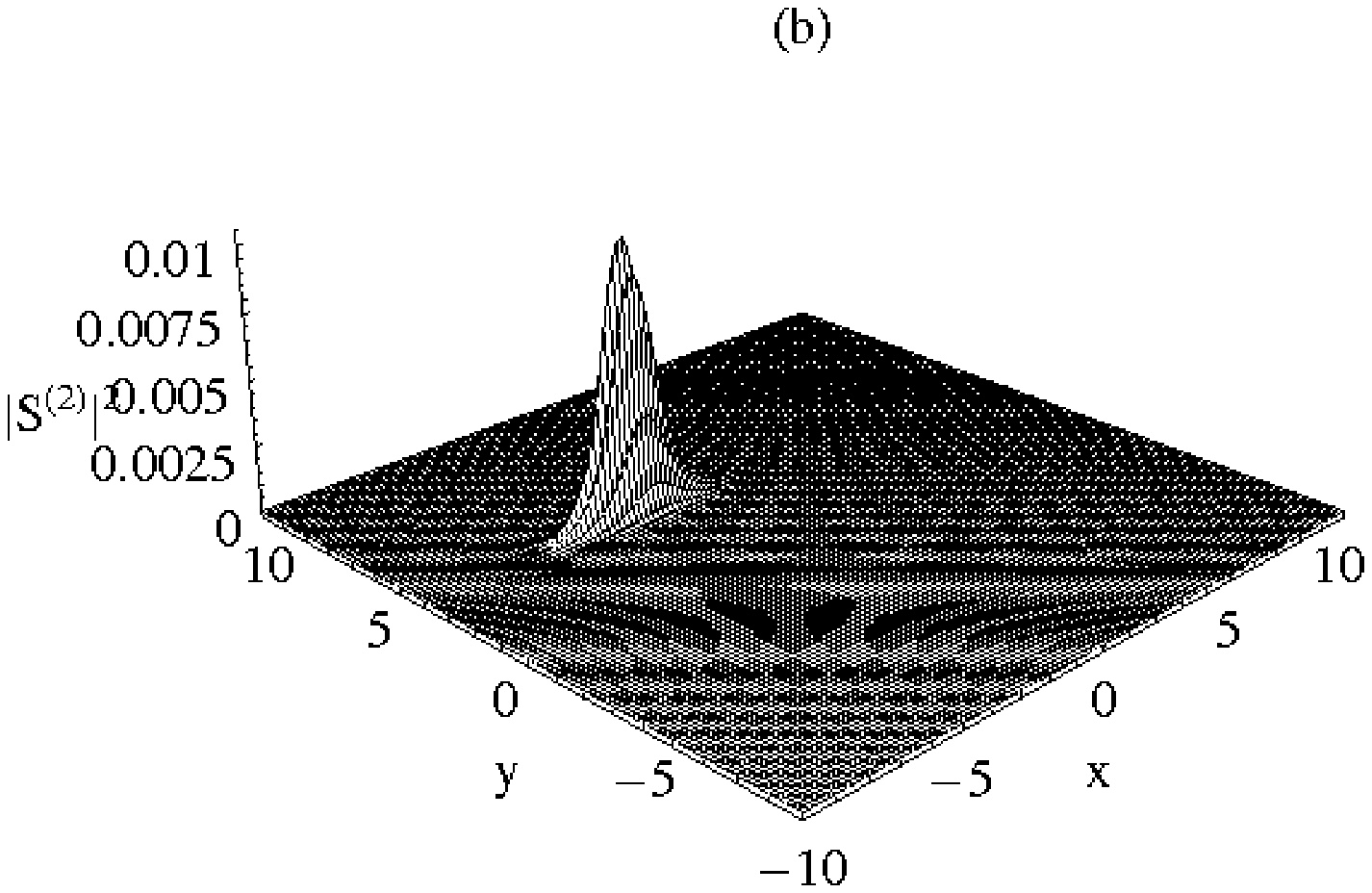, width=0.4\linewidth}
\epsfig{figure=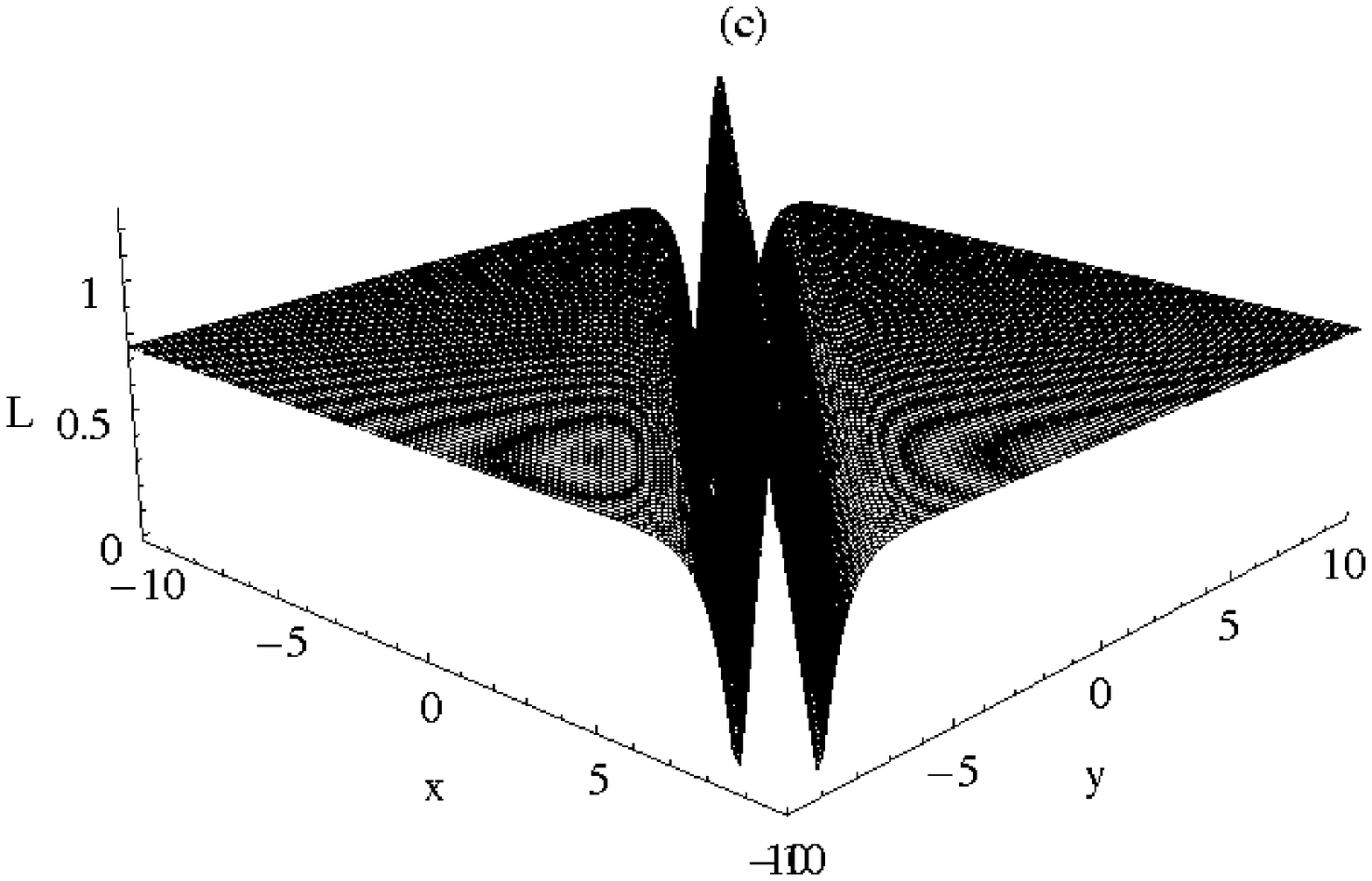, width=0.4\linewidth}
\caption{Intensity
profile of the one dromion solution for (a) the first mode, (b) the second mode, (c) line soliton for the long wave
component L at t=3.}
\end{figure}

Then, the corresponding exponentially localized solutions for
$|S^{(1)}|^2$ and $|S^{(2)}|^2$ can be written as

\begin{eqnarray}
&&\hspace{-2cm} |S^{(1)}|^2=\frac{\begin{array}{l}
 (a_1 \sec h(d_1 (t - y) + e_1 ) + g_1 )^2 {a_2 b_2 \sec h(b_2 x + c_2 y + e_2 )^2 }  \\
\end{array}}{{(a_2 \tanh (b_2 x + c_2 y + e_2 ) + g_2  + g_3  + a_4 \tanh (d_4 (t - y) + e_4 ) + g_4)^2 }}
,\nonumber\\
\nonumber\\
 &&\hspace{-2cm} |S^{(2)}|^2=\frac{\begin{array}{l}
  \begin{array}{l}
 \left[a_4 d_4 \sec h(d_4 (t - y) + e_4 )^2  - (a_1 \sec h(d_1 (t - y) + e_1 ) + g_1 )^2 \right] \times  \\
 a_2 b_2 \sec h(b_2 x + c_2 y + e_2 )^2  \\
 \end{array}   \\
  \end{array}}{{(a_2 \tanh (b_2 x + c_2 y + e_2 ) + g_2  + g_3  + a_4 \tanh (d_4 (t - y) + e_4 ) + g_4 }}.
\end{eqnarray}

The variable $L$ takes the form
\begin{eqnarray}
L&=& \frac {2 a_2^2 \sec h (b_2+c_2 y+e_2)^4}{(a_2 \tanh (b_2 x +
c_2 y + e_2 ) + g_2  + g_3  + a_4 \tanh (d_4 (t - y) + e_4 ) +
g_4)^2}\nonumber \\
& &- \frac {4 a_2 b_2^2 \sec h (b_2+c_2 y+e_2)^2 \tanh (b_2+c_2
y+e_2)}{(a_2 \tanh (b_2 x + c_2 y + e_2 ) + g_2 + g_3 + a_4 \tanh
(d_4 (t - y) + e_4 ) +
g_4)}\nonumber \\
& &  -\frac{1}{4} \frac{c_2^2}{b_2^2}+b_2^2
\end{eqnarray}

A plot of the one dromion solution for the modes $S^{(1)}$ and
$S^{(2)}$ for the following parametric choice, $a_1 = 1; a_2 = a_4
= 0.6; b_2 = 1; c_2 = 1; d_1 = d_4 = 4; e_1 = e_2 = e_4 = 0; g_1 =
g_2 = g_4 = 0; g_3 = 3 (a_4d_4>a_1^2)$ is shown in figs. 1(a) and 1(b).
From the figures, it is clear that the dromions for the modes
$S^{(1)}$ and $S^{(2)}$ moving in the y-direction have different
amplitudes and the amplitude of the dromions and hence the energy
in a given mode depends on the parameter $a_1$. We call such
exponentially localized solutions driving $S^{(1)}$ and $S^{(2)}$
as "multimode dromions". Further, the above choice of lower
dimensional arbitrary functions of space and time given by eq.
(30) yields a line soliton for the long wave L as shown in fig. 1(c).

To generate a $(2,1)$-dromion for $S^{(1)}$ and $S^{(2)}$, we
choose
\begin{eqnarray}
F_1&& = a_1 sech (d_1 (t- y) +e_1) +g_1\nonumber \\
\phi_1&& = a_2 tanh(b_2 x +c_2 y + e_2) +\nonumber\\
\quad && a_3 tanh(b_3 x +c_3 y +e_3)
+g_2\\
\phi_{21}&&= g_3, \phi_{22} = a_4 tanh(d_4 (t -y)
+e_4)+g_4\nonumber
\end{eqnarray}

so that the explicit solution can be written as
\begin{eqnarray}
&&\hspace{-2cm}|S^{(1)}|^2=\frac{\begin{array}{l}
 (a_1 \sec h(d_1 (t - y) + e_1 ) + g_1 )^2 \begin{array}{l}
 [a_2 b_2 \sec h(b_2 x + c_2 y + e_2 )^2  \\
  + a_3 b_3 \sec h(b_3 x + c_3 y + e_3 )^2] \\
 \end{array}
  \end{array}}{\begin{array}{l}
 [a_2 \tanh (b_2 x + c_2 y + e_2 ) + a_3 \tanh (b_3 x + c_3 y + e_3 ) + g_2  + g_3  +  \\
 a_4 \tanh (d_4 (t - y) + e_4 ) + g_4]^2 \\
 \end{array}},\nonumber\\
 \nonumber\\
&&\hspace{-2cm}|S^{(2)}|^2=\frac{\begin{array}{l}
  \begin{array}{l}
 \left[a_4 d_4 \sec h(d_4 (t - y) + e_4 )^2  - (a_1 \sec h(d_1 (t - y) + e_1 ) + g_1 )^2\right] \times  \\
 \left[a_2 b_2 \sec h(b_2 x + c_2 y + e_2 )^2  + a_3 b_3 \sec h(b_3 x + c_3 y + e_3 )^2\right] \\
 \end{array}
  \end{array}}{\begin{array}{l}
 [a_2 \tanh (b_2 x + c_2 y + e_2 ) + a_3 \tanh (b_3 x + c_3 y + e_3 ) + g_2  + g_3  +  \\
 a_4 \tanh (d_4 (t - y) + e_4 ) + g_4]^2  \\
 \end{array}}.
\end{eqnarray}

The plot of the (2,1) dromion solution for the modes $S^{(1)}$ and
$S^{(2)}$ for the following parametric choice $a_1 = 1; a_2 = 1;
a_3 = 1; a_4 = 1; b_2 = b_3 = 1; c_2 = 1; c_3 = -1; d_1 = d_4 = 4;
e_1 = 0; e_2 = e_3 = 0; e_4 = 0; g_1 = g_2 = g_4 = 0; g_3 =
10(a_4d_4>a_1^2)$ at t=-6, -4, -1, 5 is shown in figs. (2a-2h).

\begin{figure}
\epsfig{figure=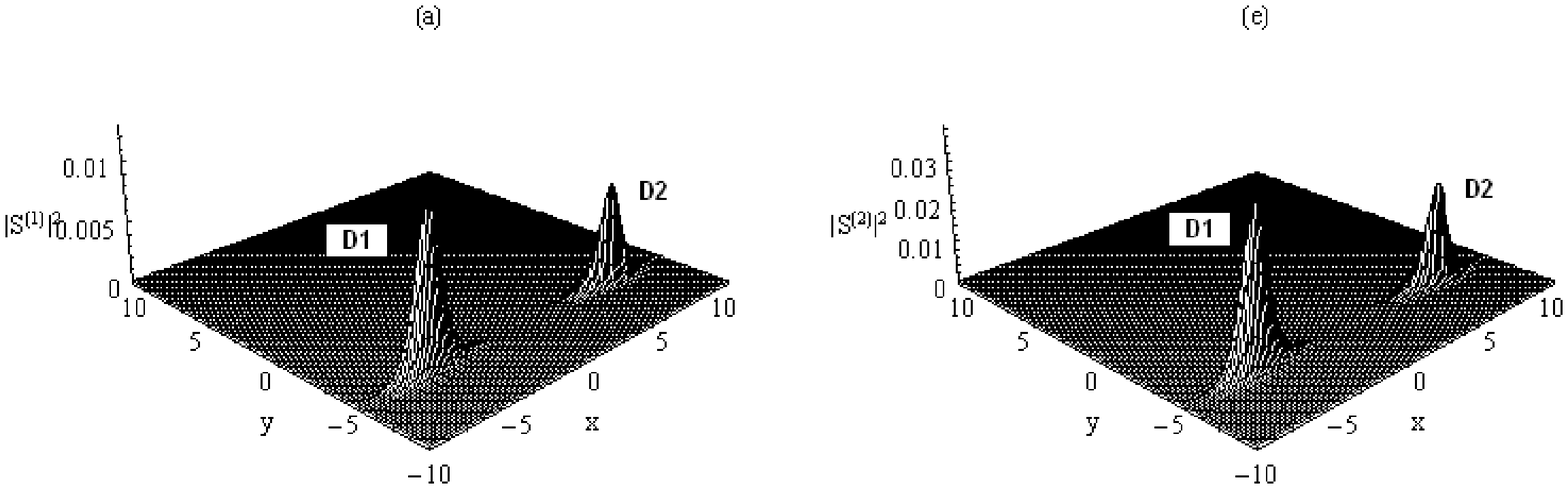, width=0.80\linewidth}
\epsfig{figure=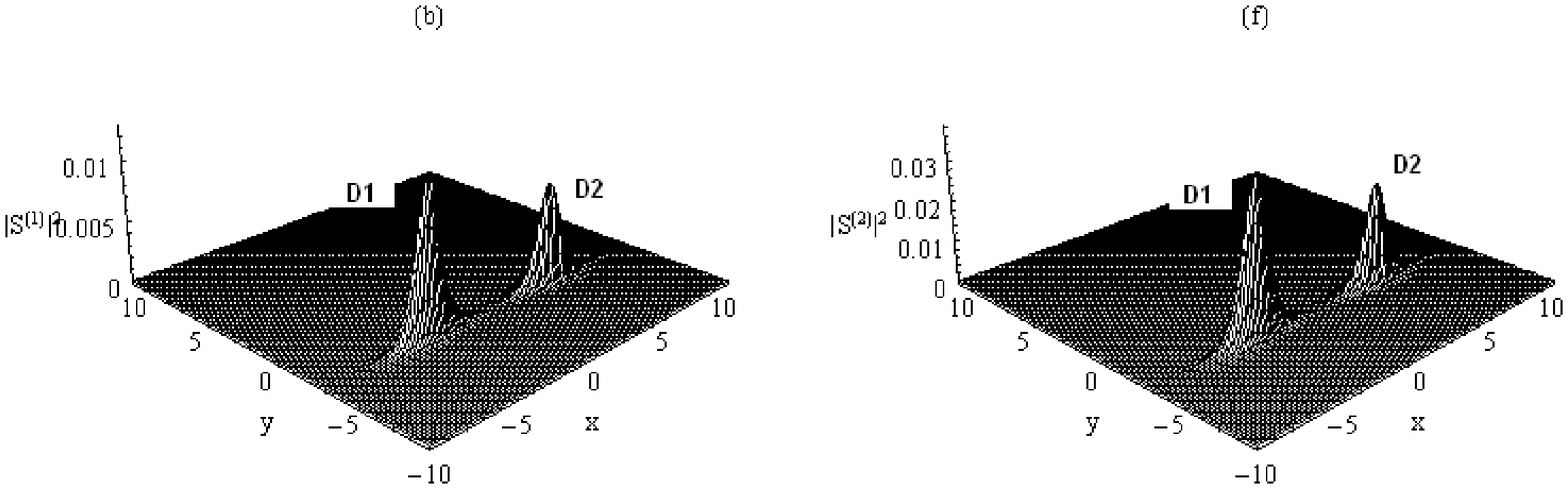, width=0.80\linewidth}
\epsfig{figure=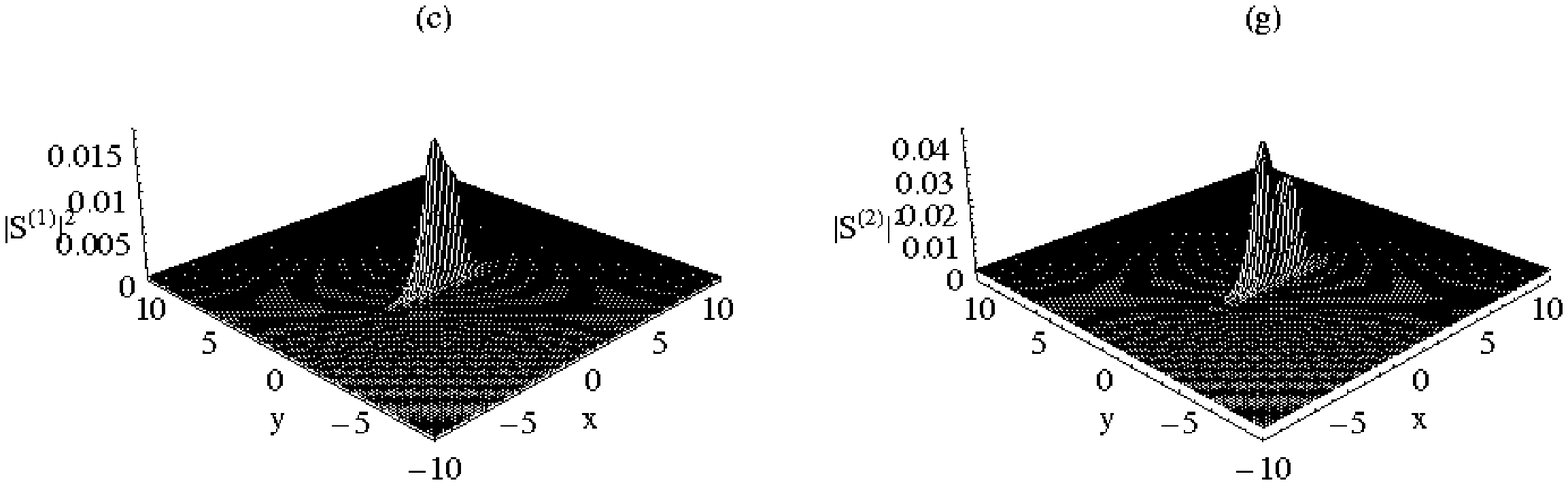, width=0.80\linewidth}
\epsfig{figure=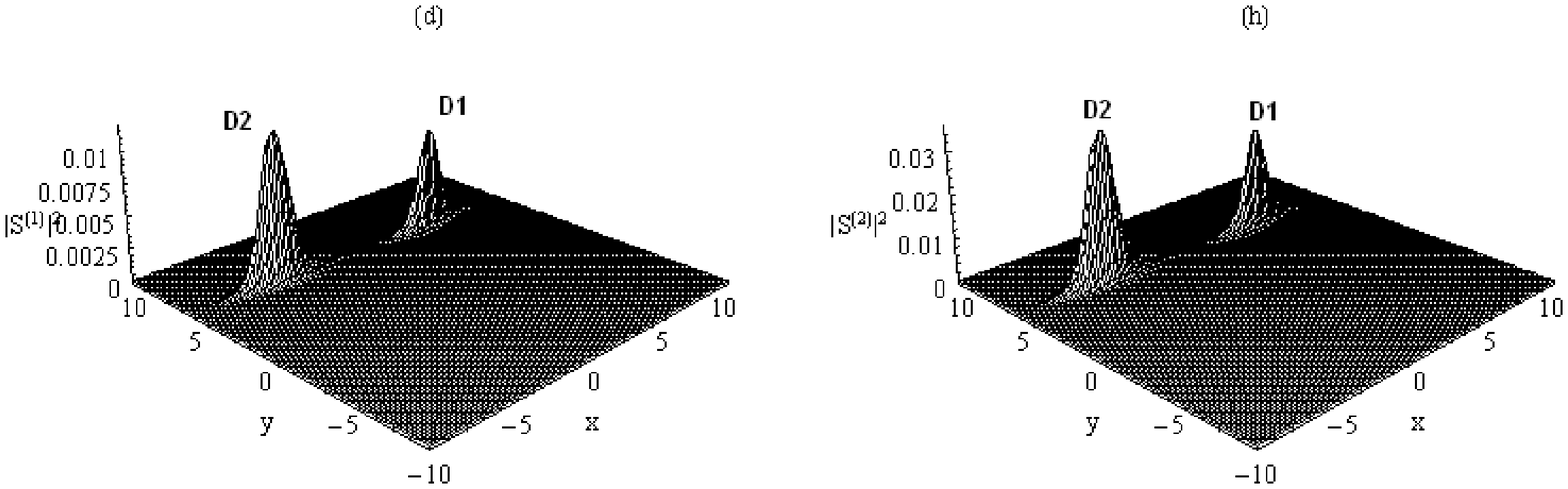, width=0.80\linewidth}
 \caption {Intensity profiles of the two dromion solution for the first mode (4a-4d)
and second mode (4e-4h) at t = -6.0, -4.0, -1.0, 5.0.}
\end{figure}

From the interaction of the dromions for the modes $S^{(1)}$ and
$S^{(2)}$ shown in figs. (2a-2h), one observes that the two
exponentially localized solutions with initial intensities $D_1$
and $D_2$ ($D_1>D_2$) move along the diagonals in the (x-y) plane
and exchange their intensities (energies) among themselves after
interaction ($D_1<D_2$) thereby undergoing intramodal inelastic
collision. It is also interesting to note that there is no
exchange of energy between the two constituent modes and the
energy contained in a given mode remains a constant.

It should be mentioned that the choice of the lower dimensional arbitrary functions of space and time $F_1(t-y)$, $\phi_1(x,y)$, $\phi_2(y)$ and
$\phi_{22}(t-y)$ determine the nature of the solutions admitted by 2CLSRI equation and their collision dynamics.
For the choice of arbitrary functions given by eq.(30), one observes that the short waves are driven by exponentially localized
solutions (dromions) and the energy contained in the first mode $S^{(1)}$ depends on $F_1(t-y)^2$ while for the second mode $S^{(2)}$, it is
governed by $(\phi_{22,t}-F_1(t-y)^2)$. For the choice given by eq.(30) (with the parameters as in fig.4), the amplitude (energy) of the first mode $S^{(1)}$
is governed by the one dimensional soliton $sech^2 4(t-y)$ while for the second mode $S^{(2)}$, it depends on the soliton $3sech^2 4(t-y)$. Thus,
the choice given by eq. (30) launches two different energies in the modes $S^{(1)}$ and $S^{(2)}$ governed by the one dimensional solitons $sech^2 4(t-y)$ and $3sech^2 4(t-y)$, respectively, and
since the amplitude (energy) of the solitons does not change during evolution, the energy contained in a mode remains a constant.
Quantitatively, this is governed by the condition
\begin{equation}
\frac{|S^{(1)}|^2}{|S^{(2)}|^2}=\frac{F_1^2}{\phi_{22,t}-F_1^2}\quad\overrightarrow{t\rightarrow\pm\infty}\quad\frac{g_1^2}{g_4^2-g_1^2}=\rm{constant}
\end{equation}
The above condition explains the existence of a firewall across the modes. This prohibition of energy across the modes by virtue of the existence of a firewall is valid only for the choice given by eq.(30), particularly if the short waves are to be driven by dromions.
This behaviour in a vector (2+1) dimensional nonlinear pde is in sharp contrast to the
Manakov model, a vector (1+1) nonlinear Schrodinger equation
wherein the energy associated with the one dimensional solitons
keeps flowing from one mode to the other. It should be mentioned
that we report for the first time the identification of
exponentially localized solutions in a vector (2+1) dimensional
nonlinear pde and their collision dynamics.

From eqns. (27a) and (27b), one also observes that the sum of the
squares of the short waves $S^{(1)}$ and $S^{(2)}$ obeys the
following equation
\begin{equation}
|S^{(1)}|^2+|S^{(2)}|^2=\frac{\phi_{22,t}\phi_{1x}}{(\phi_1(x,y)+\phi_{21}(y)+\phi_{22}(t-y))^2}=S^{(1,2)}
\end{equation}
\begin{figure}
\epsfig{figure=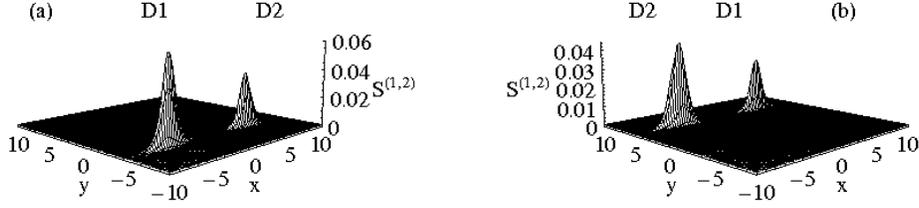, width=0.8\linewidth}\caption {Time
evolution of the composite mode $S^{(1,2)}$ at t= $-$6.0, 5.0}
\end{figure}
where we call $S^{(1,2)}$ as the "composite mode". Thus, we find
that the composite mode $S^{(1,2)}$ is again driven by a two
dromion solution (shown in figs.3((a),(b)) at t=$-$6, 5). The
intensity of the solution for the composite mode $S^{(1,2)}$ is
the sum of the constituent modes $S^{(1)}$ and $S^{(2)}$ at every
instant of time and one also observes a similar inelastic
collision in the composite mode $S^{(1,2)}$.

\section{Conclusion}
In this paper, we have investigated the two component LSRI
equation and shown that it admits Painleve property. We have then
suitably exploited the truncated Painleve approach and generated
multimode dromions for the short waves $S^{(1)}$ and $S^{(2)}$.
The collision dynamics of multimode dromions generated in the
paper indicates that they suffer intramodal inelastic collision
while the existence of a firewall prevents the flow of energy from
one mode to the other. It would be interesting to investigate the
n-component LSRI equation from the perspective of localized
solutions and their interaction.

$\textbf{\textrm{Acknowledgements:}}$ RR wishes to acknowledge the financial assistance received from DST and UGC in the form of major and minor research projects.
She also wishes to thank Indian National Science Academy (INSA) and
Royal Society of London for sponsoring her visit to Glasgow under
the bilateral exchange programme. The work of ML forms part of a
Department of Science and Technology, Govt. of India research
project and is also supported by a DST Ramanna Fellowship.

\end{document}